\documentclass[a4paper,fleqn,usenatbib]{mnras}

\usepackage{mathptmx}

\usepackage[T1]{fontenc}
\usepackage{ae,aecompl}

\usepackage{graphicx}	
\usepackage{amsmath}	
\usepackage{amssymb}	
\usepackage{color,soul}       
\usepackage{ulem}

\title[Low pressure charging]{Collisional Charging in the Low Pressure Range of Protoplanetary Disks}

\author[Becker et al.]{
T. Becker$^1$\thanks{E-mail: timsven.becker@uni-due.de}, T. Steinpilz$^1$, J. Teiser$^1$, G. Wurm$^1$\\
$^1$University of Duisburg-Essen, Faculty of Physics, Lotharstr. 1, 47057 Duisburg, Germany\\
}

\pubyear{2022}

\begin{document}
\label{firstpage}
\pagerange{\pageref{firstpage}--\pageref{lastpage}}
\maketitle

\begin{abstract}
In recent years, collisional charging has been proposed to promote the growth of pebbles in early phases of planet formation. Ambient pressure in protoplanetary disks spans a wide range from below $10^{-9}$ mbar up to way beyond mbar. Yet, experiments on collisional charging of same material surfaces have only been conducted under Earth atmospheric pressure, Martian pressure and more generally down to $10^{-2}$ mbar thus far. This work presents first pressure dependent charge measurements of same material collisions between $10^{-8}$ and $10^3$ mbar. Strong charging occurs down to the lowest pressure. In detail, our observations show a strong similarity to the pressure dependence of the breakdown voltage between two electrodes and we suggest that breakdown also determines the maximum charge on colliding grains in protoplanetary disks. We conclude that collisional charging can occur in all parts of protoplanetary disks relevant for planet formation. 
\end{abstract}


\section{Introduction} \label{sec:intro}

The motivation behind this work originates in the potential of collisional charging for planet formation. There are a number of barriers limiting the growth from dust towards larger objects. Starting at small aggregates at higher elevations of the disk a charging barrier has been proposed as grains preferentially charge negatively and Coulomb repulsion prevents collisions eventually \citep{Okuzumi2009, Okuzumi2011}. If aggregates manage to grow to millimeter size they encounter a bouncing barrier as they are now colliding mostly elastically after compaction \citep{Zsom2010, Kelling2014, Kruss2016}. Even larger aggregates would start to fragment instead of growing \citep{Birnstiel2016, Wurm2021}, noting that there are still parameter combinations that allow growth even in high speed collisions \citep{Wurm2005, Teiser2009}.
The two latter barriers strongly depend on the sticking properties of the grains. These are van der Waals forces for neutral grains. However, it has been shown in a number of works recently that tribocharging or collisional charging might sustain further growth as it adds strong electrostatic forces \citep{Lee2015, Steinpilz2020a, Teiser2021, Jungmann2021c}.

Grains do not charge infinitely though. Field induced breakdown is one of the limiting factors to how much charge a particle can hold and thus, how much charge it can gain from small scale collisions \citep{Wurm2019, Schoenau2021, Jungmann2021b, Mendez2016, Matsuyama2018}.  
So far, pressure dependent charge measurements were limited to a range from about $10^{-2}$ mbar to normal pressure \citep{Wurm2019, Mendez2018, Merrison2012}. In this range discharges occur due to atmospheric breakdown. In their experiments \cite{Wurm2019} found that below the Paschen minimum at about 1 mbar charge increases again, which still is in agreement to atmospheric breakdown. 
However it was unknown how charging would continue for ever lower pressures. This is of relevance for planet formation, as the pressure in protoplanetary disks varies strongly throughout its radial and vertical extent by orders of magnitudes. The pressure in the midplane can reach $1$ mbar in the inner parts and be as low as $10^{-9}$ mbar at 100 AU \citep{Wood2000}. 

It is not certain if charging occurs in the whole pressure range. E.g., if volatiles on grain surfaces, such as water \citep{Lee2018}, were the agents of charge transfer, low pressure might deprive the grains of potential charge carriers. 
However, monolayers of water might still prevail and be sufficient to warrant charging.
With this in mind, we did set up an experiment to measure the pressure dependency of collisional charging.

\section{Experiments}

\begin{figure}
    \centering
    \includegraphics[width= \columnwidth]{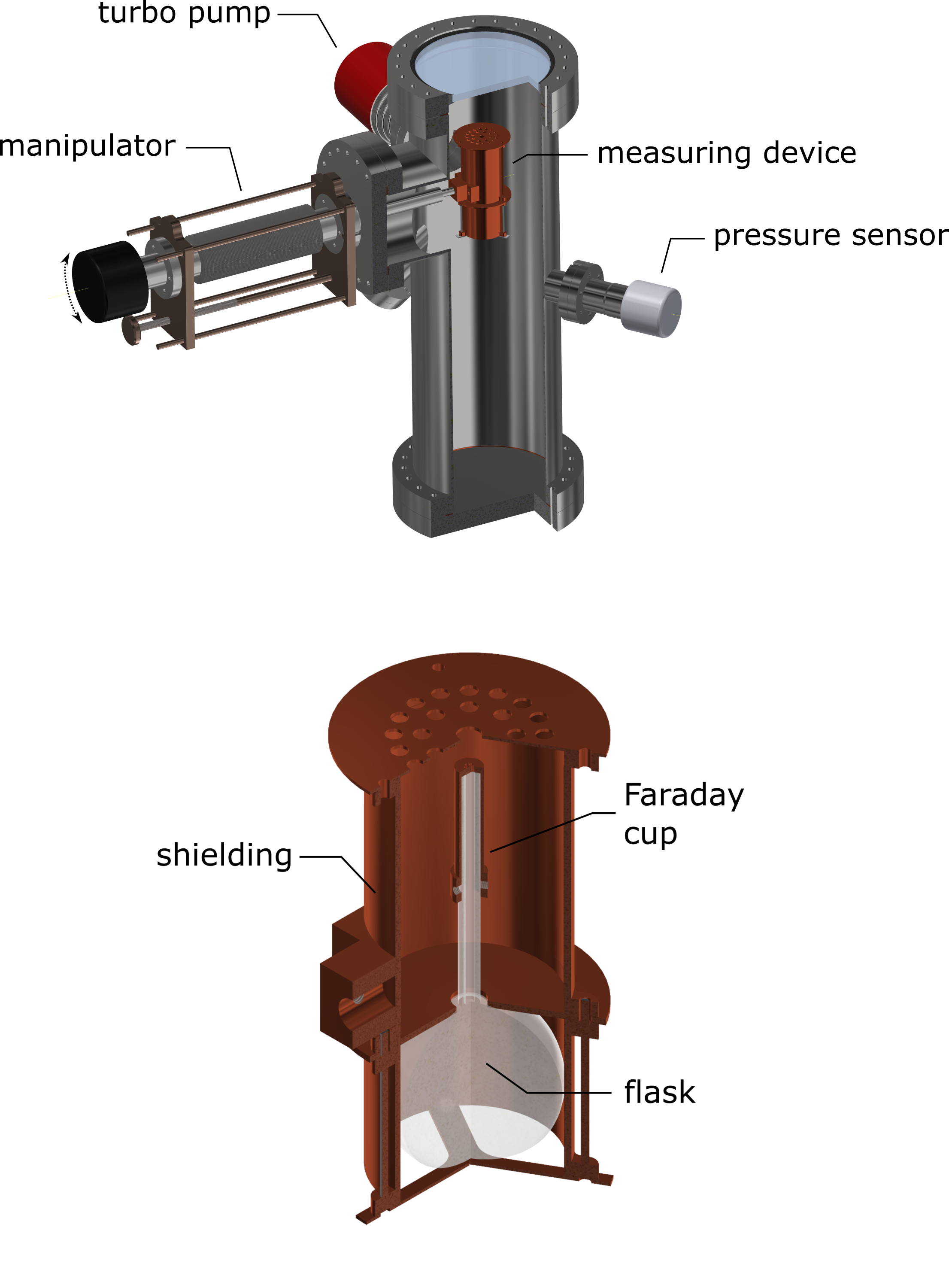}
    \caption{Sketch of the experiment. Top: Basic experiment components. The measuring device is shown in detail below. Here, the glass particle (not shown) can move freely within the glass flask and upon rotation drops into a Farday cup.}
    \label{fig:setup}
\end{figure}

The experiment is sketched in fig. \ref{fig:setup}. The method is rather straightforward. A spherical glass particle collides with a glass surface and its charge is measured afterwards by means of a Faraday cup.
The charging setup is placed in a vacuum chamber so that the charge can be measured at pressures of $1000$ mbar down to $10^{-8}$ mbar. 

\begin{figure}
    \centering
    \includegraphics[width= \columnwidth]{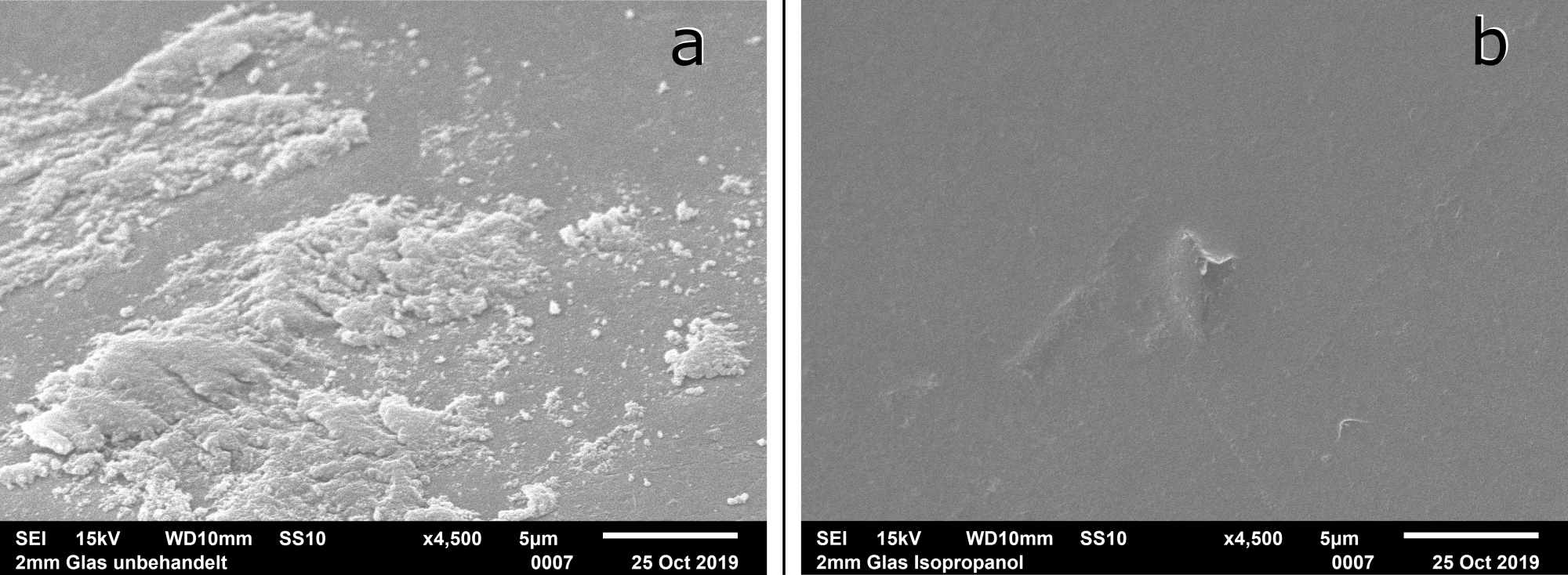}
    \caption{SEM images of \textit{a)} an untreated particle surface and \textit{b)} a surface irregularity after treatment with isopropanol. Note that most of the surface of the particle is clean after treatment but we did chose an image to highlight the small irregularities that might be left.}
    \label{fig:surface}
\end{figure}

We used borosilicate glass for the particle and its counterpart. We had both surfaces made of the same material in order to prevent a material dependent bias in tribocharging. Additionally, we paid attention to thoroughly clean all surfaces. For all glass surfaces an isopropanol ultrasonic bath was used. It turned out to deliver the best results, compared to other cleaning methods. Fig. \ref{fig:surface} shows a comparison of an SEM image of an untreated surface and a surface cleaned with isopropanol.
While on the untreated surface major contamination is visible, the cleaned surface only shows what appears to be a dent.
The spherical partile is 2 mm in diameter. Its counter surface is a balloon-shaped flask with an inner diameter of $48$ mm. On the flask´s neck a Faraday cup is mounted. The particle charges through interaction with the surface of the glass flask. 

For every measurement the assembly in which the glass flask is fixated was shaken using the manipulator it was mounted on. Then it would be rotated by $180^\circ$ for the bead to drop into the Faraday cup, measuring its charge. Afterwards it was rotated back into its initial position, allowing the bead to drop back into the flask. For every pressure this procedure was repeated 10-25 times, depending on the variation of measured values. 
Measurements have been conducted from $1000$ mbar to $10^{-8}$ mbar. The pressure was held constant during all measurements.

\section{Results and Discussion}

\begin{figure}
    \centering
    \includegraphics[width= \columnwidth]{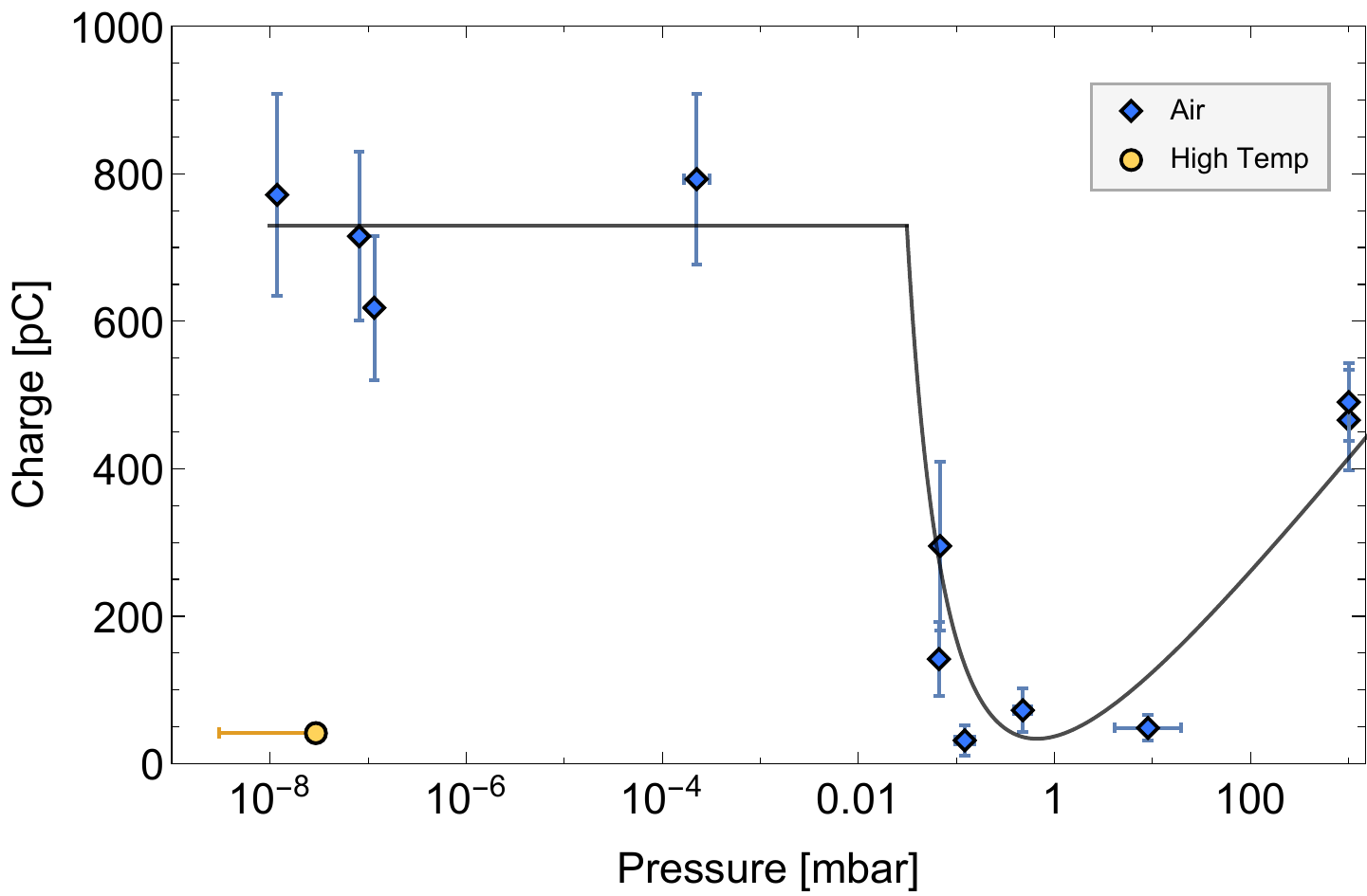}
    \includegraphics[width= \columnwidth]{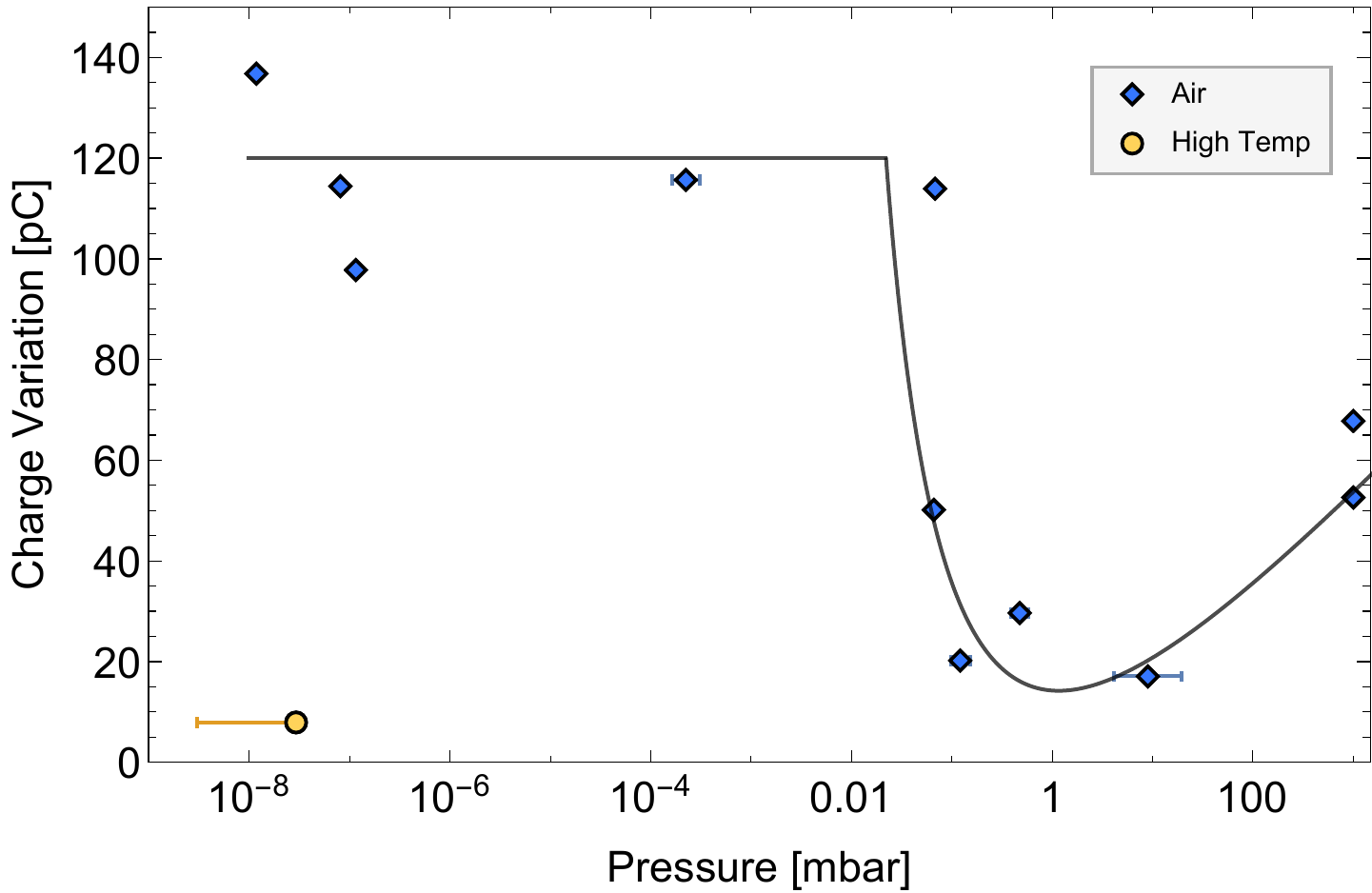}
    \caption{Pressure dependence of the net charge (top) and charge variation (bottom) on a tribocharged glass particle in air. Overplotted is a Paschen curve above $0.02$ mbar with constant distance (see text for details). Below that is a constant fit.}
    \label{fig:air}
\end{figure}

\begin{figure}
    \centering
    \includegraphics[width= \columnwidth]{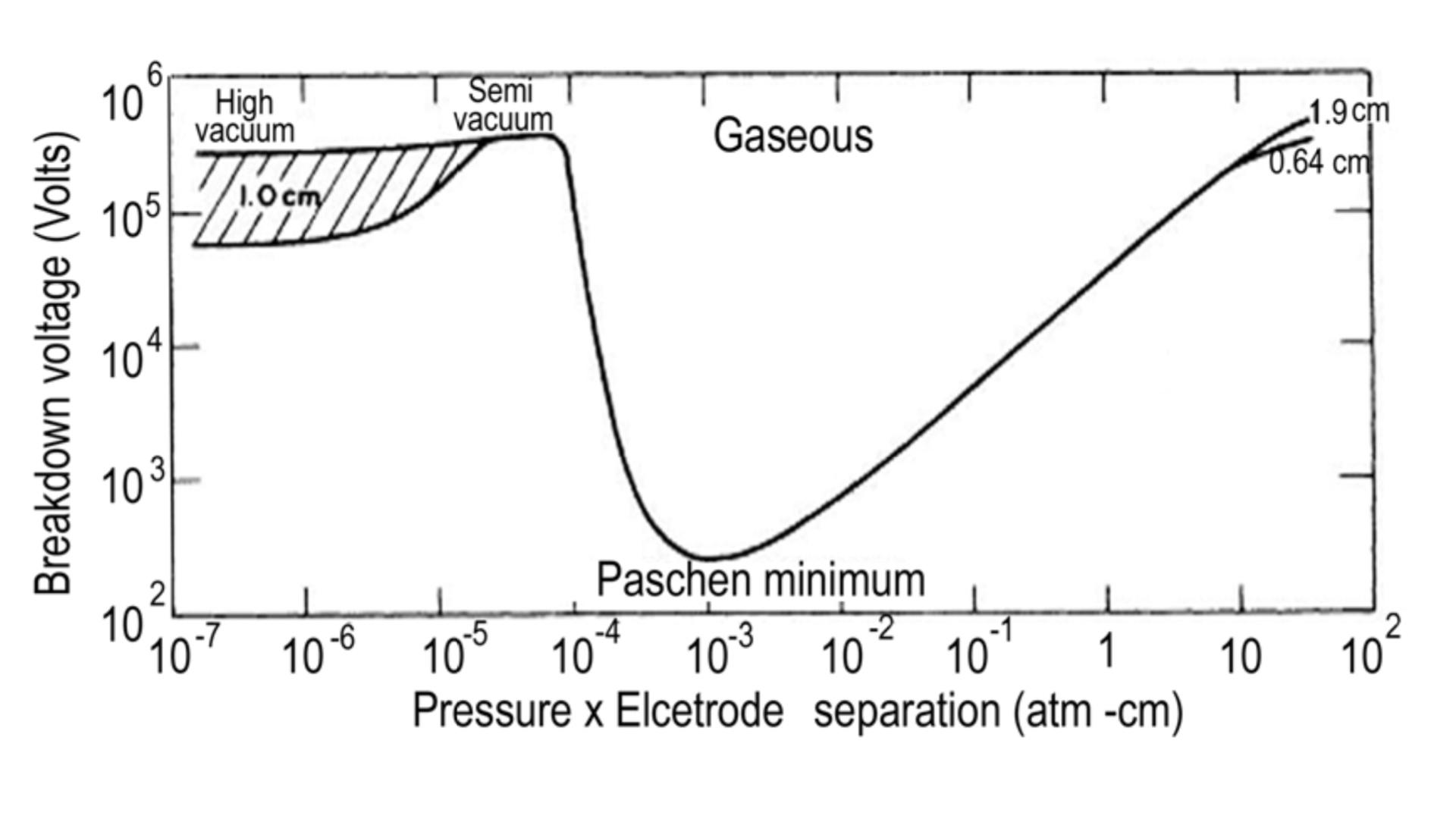}
    \caption{General scheme of breakdown voltages between two electrodes taken from \citet{Zheng2013b}}
    \label{fig:model}
\end{figure}

The measured net charge on the grain and its standard deviation over pressure is shown in fig. \ref{fig:air} (top). It can be seen that the charge decreases from atmospheric pressures to a minimum at about $1$ mbar and then increases towards a constant value down to $10^{-8}$ mbar.
If charging of the glass bead is limited to the point at which the voltage between bead and flask surpasses the atmospheric breakdown value, its charge should follow the behaviour of breakdown voltages observed in technical applications between two electrodes. The latter breakdown voltage over the entire pressure range is shown in fig. \ref{fig:model} \citep{Zheng2013b, Bommottet2013}.
As can be seen, the functional behaviour of our measurements is well represented by the breakdown curve. 

While the comparison between fig. \ref{fig:air} (top) and fig. \ref{fig:model} is striking, it has to be kept in mind that the quantities on the y-axis are different. We measure total net charge while the breakdown depends on voltage (and electrode distance). Therefore, the interpretation of the functional pressure behaviour as atmospheric breakdown has to make a connection between charge and voltage, e.g. explicitly done by \citet{Wurm2019} or \citet{Matsuyama2018}. In detail, discharge occurs as a grain approaches or retracts from the surface and its charge and the distance at which discharge occurs sets itself as soon as breakdown conditions are met. The functional behavior or the only pressure dependent discharge curve looks similar to a Paschen curve with constant distance then \citep{Wurm2019}. For simplicity, as this is mostly for visual support here, we only added an ordinary Paschen curve at the high pressure range in fig. \ref{fig:air} assuming a constant distance between electrodes and slightly shifting it in pressure to match the data points best.

In this context, it has to be considered that the net charge we measure might hold a bias.
The voltage between the bead and the glass surface depends on the charge on the glass surface. Thus, the breakdown conditions vary accordingly.
This is in analogy to measurements of \citet{Matsusaka2008}, where particles passing through tubes of different voltages acquired different net charges as well.
For the particle ensemble discussed by \citet{Wurm2019}, the average net charge of grains was zero, showing that not the total charge, but its variation holds the breakdown conditions.
In other words, if little voltage is needed for breakdown, there cannot be much variation in charge. Consequently, if the voltage for breakdown is high, so is the range for possible charge values. Variations also imply that charge transfer is taking place and that we do not measure fixed charges on or within the grain. Therefore, we also give the standard deviations over pressure in fig. \ref{fig:air} (bottom).
In our case, the figure shows the same qualitative behaviour as the top panel except for the total net charges being smaller. 

Our data suggest breakdown to be the limiting factor. Classically, breakdown has two regimes. That is the high pressure range beyond $\sim 10^{-2}$ mbar, where breakdown can be described by the Paschen-curve. Free electrons trigger further ionization of the gas, which leads to a Townsend avalanche and thus conduction or discharge between electrodes.
As detailed in \citet{Wurm2019} for the higher pressure end, this is also true for particle-particle collisions. It requires the particle and surface to separate by a distance larger than the mean free path of the molecules after a collision \citep{Schoenau2021}. If that is achieved, a similar pressure profile for charge as given by the Paschen-curve for voltages follows.
Below $\sim 10^{-2}$ mbar, the mean free path is of the order of the glass flask´s inner diameter or larger and thus the ambient gas can no longer support breakdown, no matter, where the grain is exactly located within the flask. Also between two electrodes, breakdown is no longer considered to be related to the residual gas if the mean free path is larger than the electrode distance \citep{Zheng2013b}. 
Instead, field emission and vaporization of electrode surfaces is assumed to trigger discharge in this regime \citep{Zheng2013b}. The similarity of our charge data to breakdown voltages suggests that something similar is occurring for tribocharged grains.

\subsection{High temperature measurement}

Our measurements show, that charging occurs even at the lowest pressures, meaning there are still free charge carriers available. While we do not know the source of those charge carriers, changing the environment might bring some insight into the matter. To see if a change in temperature has an effect on the charging behaviour,
we conducted measurements at $388$K and $10^{-8}$ mbar after heating the vacuum chamber for three days at $388$K.
The absolute charges measured, as well as variations between measurements were extremely low compared to other measurements at $10^{-8}$ mbar as can be seen in fig. \ref{fig:air}.  
The dramatic decrease of charge at high temperature means that either electrification is reduced, discharge is enhanced, or a combination of the two effects. While we cannot be certain,
low net charge and low variability is well in agreement to a lack of charge carriers. 
Given the net charge available and transferable is minuscule, an electric field for field emission cannot build up.
However, as it is just one data point, further speculation would be pointless. Thus, for now, we just take it as a finding that increased temperature seems to have a negative effect on collisional charging at low pressures for our setting of glass as sample.

\section{Charging in protoplanetary disks}   

Our results show that collisional charging is possible throughout all regions of protoplanetary disks that are relevant for planet formation. While we cannot determine which processes are responsible for charging or discharging in detail, we find that low pressure down to $10^{-8}$ mbar does not restrict charging. In fact, it rather has the opposite effect. 

The measurement at high temperature implies that water is important for charge transfer. 
In protoplanetary disks, water is abundant. Additionally, pressure and temperature usually depend on the radial distance from the star, meaning, low pressure comes with low temperatures and high pressure with high temperatures. This work neither gives any results on charging in a cold environment, where ice particles dominate as species, nor for high temperatures at high pressures. However, for the latter, \citet{Mendez2018} showed, that collisional charging also occurs in this regime.


\section{Conclusion}

We measured the pressure dependence of the net charge on a spherical glass bead colliding with another, basically flat glass surface. The maximum charges are qualitatively in agreement to breakdown theory, explaining a high charge at very low pressure and a minimum charge at intermediate pressure. It is likely that the charging is moderated by water ions. We found that charging is possible at all pressures studied in this work and therefore collisional charging can occur - at least with our parameters - in all parts of protoplanetary disks relevant for planet formation.
We conclude that collisional charging may be a significant process in planet formation.

\section*{acknowledgments}
This project has received funding from the European
Union’s Horizon 2020 research and innovation
program under grant agreement No 101004052. We thank Jonathan Merrison for reviewing this paper.

\section*{Data Availability}
The data underlying this article will be shared on reasonable request to the corresponding author.

\bibliography{bib}{}
\bibliographystyle{mnras}

\bsp	
\label{lastpage}
\end{document}